# The Driven Liouville von Neumann Equation in Lindblad Form

Oded Hod,[1] César A. Rodríguez-Rosario,[2] Tamar Zelovich,[1] Thomas Frauenheim[2]

*1) Department of Physical Chemistry, School of Chemistry, The Raymond and Beverly Sackler Faculty of Exact Sciences and The Sackler Center for Computational Molecular and Materials Science, Tel Aviv University, Tel Aviv 6997801, Israel*

*2) Bremen Center for Computational Materials Science, University of Bremen, Am Falturm 1, Bremen, 28359, Germany.*

**Abstract:**

The Driven Liouville von Neumann approach [*J. Chem. Theory Comput.* **10**, 2927-2941 (2014)] is a computationally efficient simulation method for modeling electron dynamics in molecular electronics junctions. Previous numerical simulations have shown that the method can reproduce the exact single-particle dynamics while avoiding density matrix positivity violation found in earlier implementations. In this study we prove that, in the limit of infinite lead models, the underlying equation of motion can be cast in Lindblad form. This provides a formal justification for the positivity and trace preservation obtained numerically.

## Introduction

Since its inception over 40 years ago[1] the field of molecular electronics holds the promise to use individual molecular entities and their assemblies as active components in electronic devices.[2-4] Molecules are often characterized by their miniature size, quantum mechanical nature, well defined chemical composition, highly efficient and accurate synthesis procedures, as well as self-assembly capabilities. These properties open the door for the design of novel molecular scale electronic components that may present unique functionality with high sensitivity towards external perturbations and fast response time. Furthermore, such systems are expected to be energetically efficient and to allow for cost-effective reproducible device fabrication.

Steady-state transport properties of molecular junctions remain the main focus of both experimental and theoretical efforts in this field. Nevertheless, the study of dynamical transport phenomena in nanoscale junctions has recently gained increasing attention from the scientific community.[5-32] Research in this direction explores the effects of time-dependent perturbations such as alternating currents, bias pulses, and external electromagnetic fields on the transient response of the system. This involves complex physical processes that can be harnessed for the design of miniaturized electronic devices such as opto-electronic ultrafast molecular switches and nanoscale rectifiers.[7,8,14,22,24-27,29,31]

Considering the temporal degree of freedom poses new challenges on the experimental efforts that theory and simulation may help resolve via the prediction of optimal junction configurations and operation conditions and the interpretation of experimental findings. To this end, various methods aiming to simulate electron dynamics in molecular junctions have been developed.[31-89] These approaches either consider the detailed atomistic structure of the junction or introduce model Hamiltonians to study specific transport phenomena. Furthermore, they vary in the level of treatment of electron-electron interactions and effects of bath memory.

Recently, we proposed the Driven Liouville von Neumann approach that imposes dynamic non-equilibrium boundary conditions on finite atomistic models of realistic molecular junctions.[90,91] This method was shown to provide a decent compromise between computational efficiency and physical accuracy when modeling single-electron dynamics in molecular junctions subjected to time-dependent external perturbations.[92]

Notably, the underlying equation of motion was numerically shown to solve the density matrix positivity conservation violation found in earlier related implementations.[80] Nevertheless, no rigorous proof was offered to support the general validity of the positivity condition for this method. In the present study, we show that in the limit of infinite lead models the Driven Liouville von Neumann equation can be written in Lindblad form. The latter guarantees preservation of density matrix positivity throughout the dynamics[93-95] thus justifying the numerical findings.

## **Derivation**

We consider the system depicted in Fig. 1. Here, a molecule bridges the gap between two leads that are connected to two external baths. We formally divide the system into five sections including the left and right semi-infinite reservoirs, the left and right leads, and the extended molecule region, which is the molecule itself augmented by its adjacent lead sections. The size of these lead sections is chosen such that the electronic properties of the full extended molecule converge to within a desired accuracy. In the present derivation, we do not treat the bath sections explicitly and instead we take them into account implicitly by considering their effect on their adjacent lead sections. So, in practice, we consider a system constructed from the finite left and right leads models connected by the extended molecule. The corresponding sets of orthonormal eigenstates of each individual section are marked as $\{|l\rangle\}$, $\{|r\rangle\}$, and $\{|m\rangle\}$, respectively.

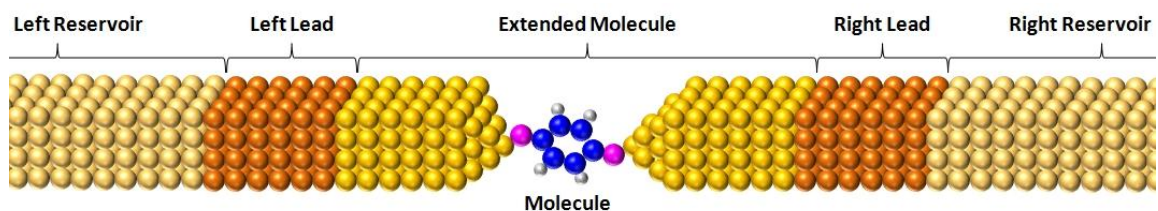

**Figure 1**: Real-space representation of a typical molecular junction model divided into the extended molecule, left and right leads, and external reservoirs.

We aim at showing that the driven Liouville von-Neumann equation that was previously derived both heuristically[90] and explicitly[92] can be written in Lindblad form thus proving that the corresponding non-unitary evolution of the density matrix is indeed trace-preserving and completely positive. To this end, we start from the Lindblad super operator that has the following general form:

$$\mathcal{L}\{\hat{\rho}\} = \sum_\alpha \left[ \hat{L}_\alpha \hat{\rho} \hat{L}_\alpha^\dagger - \tfrac{1}{2} \hat{L}_\alpha^\dagger \hat{L}_\alpha \hat{\rho} - \tfrac{1}{2} \hat{\rho} \hat{L}_\alpha^\dagger \hat{L}_\alpha \right], \tag{1}$$

where $\hat{\rho}$ is the density operator and $\hat{L}_\alpha$ are Lindblad operators that we choose in the form previously suggested by Dubi and Di-Ventra:[96]

$$\hat{L}_{k,k'} = \sqrt{\gamma_{k,k'} f_K(E_k, \mu_K)} |k\rangle\langle k'|. \tag{2}$$

Here, $|k\rangle \in \{|l\rangle\}$ or $\{|r\rangle\}$ are single-particle states in the left ($K = L$) or right ($K = R$) non-interacting lead sections, respectively. The action of these operators is to exchange charge between each finite lead model levels with (real) transition rates $\{\gamma_{k,k'}\}$ so as to maintain, as close as possible, their Fermi-Dirac equilibrium distribution $f_K(E, \mu_K) = \left[ 1 + e^{\frac{E - \mu_K}{k_B T_K}} \right]^{-1}$ induced by their implicit coupling to the corresponding reservoirs. Here, $T_K$ and $\mu_K$ are the electronic temperature and chemical potential of reservoir $K$ and $k_B$ is Boltzmann's constant. With this choice the Lindblad superoperator can be split into the individual contributions of the left and right leads $\mathcal{L}\{\hat{\rho}\} = \mathcal{L}_L\{\hat{\rho}\} + \mathcal{L}_R\{\hat{\rho}\}$ with

$$\mathcal{L}_K\{\hat{\rho}\} = \sum_{k=1}^{N_K} \sum_{k'=1, k' \neq k}^{N_K} \left[ \hat{L}_{k,k'} \hat{\rho} \hat{L}_{k,k'}^\dagger - \tfrac{1}{2} \hat{L}_{k,k'}^\dagger \hat{L}_{k,k'} \hat{\rho} - \tfrac{1}{2} \hat{\rho} \hat{L}_{k,k'}^\dagger \hat{L}_{k,k'} \right], \tag{3}$$

where the sums run over all $N_K$ levels of lead K with $k' \neq k$. In order to obtain the driven Liouville von Neumann equation we write this superoperator in matrix form. Using Eq. (2), the first term in the double sum can be written as:

$$\hat{L}_{k,k'} \hat{\rho} \hat{L}_{k,k'}^\dagger = \left[ \sqrt{\gamma_{k,k'} f_K(E_k, \mu_K)} |k\rangle\langle k'| \right] \hat{\rho} \left[ |k'\rangle\langle k| \sqrt{\gamma_{k,k'} f_K(E_k, \mu_K)} \right] = $$
$$\gamma_{k,k'} f_K(E_k, \mu_K) \rho_{k',k'} |k\rangle\langle k| \tag{4}$$

where we have denoted the matrix element $\rho_{k',k'} \equiv \langle k' | \hat{\rho} | k' \rangle$. The second term is given by:

$$-\tfrac{1}{2} \hat{L}_{k,k'}^\dagger \hat{L}_{k,k'} \hat{\rho} = -\tfrac{1}{2} \left[ \sqrt{\gamma_{k,k'} f_K(E_k, \mu_K)} |k'\rangle\langle k| \right] \left[ |k\rangle\langle k'| \sqrt{\gamma_{k,k'} f_K(E_k, \mu_K)} \right] \hat{\rho} = $$
$$-\tfrac{1}{2} \gamma_{k,k'} f_K(E_k, \mu_K) |k'\rangle\langle k'| \hat{\rho}\hat{I} = -\tfrac{1}{2} \gamma_{k,k'} f_K(E_k, \mu_K) \sum_{n \in \{|l\rangle\}, \{|m\rangle\}, \{|r\rangle\}} \rho_{k',n} |k'\rangle\langle n|, \tag{5}$$

where we introduced the identity operator within the subspace of system states: $\hat{I} = \sum_{n \in \{|l\rangle\}, \{|m\rangle\}, \{|r\rangle\}} |n\rangle\langle n|$. Accordingly, the third term can be written as:

$$-\frac{1}{2}\hat{\rho}\hat{L}_{k,k'}^{\dagger}\hat{L}_{k,k'} = -\frac{1}{2}\gamma_{k,k'}f_K(E_k, \mu_K)\sum_{n \in \{|l\rangle\}, \{|m\rangle\}, \{|r\rangle\}} \rho_{n,k'}|n\rangle\langle k'| \qquad (6)$$

Substituting Eqs. (4), (5), and (6) in Eq. (3) we obtain an expression for the Lindblad superoperator of lead K:

$$\mathcal{L}_K\{\boldsymbol{\rho}\} = \sum_{k=1}^{N_K}\sum_{k'=1}^{N_K}\gamma_{k,k'}f_K(E_k, \mu_K)\left[\rho_{k',k'}|k\rangle\langle k| - \frac{1}{2}\sum_{n \in \{|l\rangle\}, \{|m\rangle\}, \{|r\rangle\}}(\rho_{k',n}|k'\rangle\langle n| + \rho_{n,k'}|n\rangle\langle k'|)\right] - \sum_{k=1}^{N_K}\gamma_{k,k}f_K(E_k, \mu_K)\left[\rho_{k,k}|k\rangle\langle k| - \frac{1}{2}\sum_{n \in \{|l\rangle\}, \{|m\rangle\}, \{|r\rangle\}}(\rho_{k,n}|k\rangle\langle n| + \rho_{n,k}|n\rangle\langle k|)\right] \qquad (7)$$

Where we have added and subtracted the terms with $k' = k$. The eigenstates of each section ($\{|l\rangle\}$, $\{|m\rangle\}, \{|r\rangle\}$) form an orthogonal set within themselves by construction. Furthermore, since we are using a tight-binding approximation, the overlap between states of different sections is neglected thus making them mutually orthogonal. Therefore, the diagonal elements of the matrix representation of $\mathcal{L}_K\{\hat{\rho}\}$ are given by:

$$\langle i|\mathcal{L}_K|i\rangle = \sum_{k=1}^{N_K}\sum_{k'=1}^{N_K}\gamma_{k,k'}f_K(E_k, \mu_K)\left[\rho_{k',k'}\delta_{ik}\delta_{ki} - \frac{1}{2}\sum_{n \in \{|l\rangle\}, \{|m\rangle\}, \{|r\rangle\}}(\rho_{k',n}\delta_{ik'}\delta_{ni} + \rho_{n,k'}\delta_{in}\delta_{k'i})\right] - \sum_{k=1}^{N_K}\gamma_{k,k}f_K(E_k, \mu_K)\left[\rho_{k,k}\delta_{ik}\delta_{ki} - \frac{1}{2}\sum_{n \in \{|l\rangle\}, \{|m\rangle\}, \{|r\rangle\}}(\rho_{k,n}\delta_{ik}\delta_{ni} + \rho_{n,k}\delta_{in}\delta_{ki})\right]. \qquad (8)$$

Clearly, if $|i\rangle$ does not belong to the set of eigenstates $\{|k\rangle\}$ of lead K this term completely vanishes whereas when $|i\rangle \in \{|k\rangle\}$ the second sum vanishes and the only terms that survive are:

$$\langle i|\mathcal{L}_K|i\rangle = \sum_{k=1}^{N_K}\gamma_{i,k}f_K(E_i, \mu_K)\rho_{k,k} - \sum_{k=1}^{N_K}\gamma_{k,i}f_K(E_k, \mu_K)\rho_{i,i}, \qquad (9)$$

where, for brevity, we have replaced the summation index $k'$ by $k$ in the first sum.

For the off-diagonal matrix elements of $\mathcal{L}_K\{\hat{\rho}\}$ we obtain:

$$\langle i|\mathcal{L}_K|j \neq i\rangle = -\frac{1}{2}\sum_{k=1}^{N_K}\sum_{k'=1}^{N_K}\gamma_{k,k'}f_K(E_k, \mu_K)\sum_{n \in \{|l\rangle\}, \{|m\rangle\}, \{|r\rangle\}}(\rho_{k',n}\delta_{ik'}\delta_{nj} + \rho_{n,k'}\delta_{in}\delta_{k'j}) + \frac{1}{2}\sum_{k=1}^{N_K}\gamma_{k,k}f_K(E_k, \mu_K)\sum_{n \in \{|l\rangle\}, \{|m\rangle\}, \{|r\rangle\}}(\rho_{k,n}\delta_{ik}\delta_{nj} + \rho_{n,k}\delta_{in}\delta_{kj}), \qquad (10)$$

where terms involving $\delta_{ik}\delta_{kj}$ have been omitted. When $|i\rangle, |j\rangle \in \{|k\rangle\}$ this expression yields:

$$\langle i|\mathcal{L}_K|j\rangle = \frac{1}{2}\Big[\gamma_{i,i}f_K(E_i,\mu_K) + \gamma_{j,j}f_K(E_j,\mu_K) - \sum_{k=1}^{N_K}\big(\gamma_{k,i}+\gamma_{k,j}\big)f_K(E_k,\mu_K)\Big]\rho_{i,j}. \tag{11}$$

For $|i\rangle \in \{|k\rangle\}$ and $|j\rangle \notin \{|k\rangle\}$ this term gives:

$$\langle i|\mathcal{L}_K|j\rangle = -\frac{1}{2}\Big[\sum_{k\neq i}^{N_K}\gamma_{k,i}f_K(E_k,\mu_K)\Big]\rho_{i,j}. \tag{12}$$

Similarly, for $|i\rangle \notin \{|k\rangle\}$ and $|j\rangle \in \{|k\rangle\}$ we obtain:

$$\langle i|\mathcal{L}_K|j\rangle = -\frac{1}{2}\Big[\sum_{k\neq j}^{N_K}\gamma_{k,j}f_K(E_k,\mu_K)\Big]\rho_{i,j}, \tag{13}$$

And when $|i\rangle, |j\rangle \notin \{|k\rangle\}$

$$\langle i|\mathcal{L}_K|j\rangle = 0 \tag{14}$$

Collecting all the terms appearing in Eqs. (9), (11), (12), (13), and (14) we obtain the following expression for the matrix elements of the Lindblad super operator defined by Eqs. (2) and (3) in the basis of single-particle states of the separate system sections (left and right leads and the extended molecule):

$$\langle i|\mathcal{L}_K|j\rangle =$$
$$\begin{cases} \sum_{k=1}^{N_K}\gamma_{i,k}f_K(E_i,\mu_K)\rho_{k,k} - \sum_{k=1}^{N_K}\gamma_{k,i}f_K(E_k,\mu_K)\rho_{i,i} & i=j;\ |i\rangle,|j\rangle \in \{|k\rangle\} \\ 0 & i=j;\ |i\rangle,|j\rangle \notin \{|k\rangle\} \\ \frac{1}{2}\Big[\gamma_{i,i}f_K(E_i,\mu_K) + \gamma_{j,j}f_K(E_j,\mu_K) - \sum_{k=1}^{N_K}\big(\gamma_{k,i}+\gamma_{k,j}\big)f_K(E_k,\mu_K)\Big]\rho_{i,j} & i\neq j;\ |i\rangle,|j\rangle \in \{|k\rangle\} \\ -\frac{1}{2}\Big[\sum_{k\neq i}^{N_K}\gamma_{k,i}f_K(E_k,\mu_K)\Big]\rho_{i,j} & i\neq j;\ |i\rangle \in \{|k\rangle\}, |j\rangle \notin \{|k\rangle\} \\ -\frac{1}{2}\Big[\sum_{k\neq j}^{N_K}\gamma_{k,j}f_K(E_k,\mu_K)\Big]\rho_{i,j} & i\neq j;\ |i\rangle \notin \{|k\rangle\}, |j\rangle \in \{|k\rangle\} \\ 0 & i\neq j;\ |i\rangle,|j\rangle \notin \{|k\rangle\} \end{cases} \tag{15}$$

At this point we assume that all inter-state transition rates are equal and constant such that $\gamma_{k,k'} = \gamma_K = \text{Const}$ giving:

$$\langle i|\mathcal{L}_K|j\rangle =$$
$$\gamma_K \begin{cases} f_K(E_i,\mu_K)\sum_{k=1}^{N_K}\rho_{k,k} - \Big[\sum_{k=1}^{N_K}f_K(E_k,\mu_K)\Big]\rho_{i,i} & i=j;\ |i\rangle,|j\rangle \in \{|k\rangle\} \\ 0 & i=j;\ |i\rangle,|j\rangle \notin \{|k\rangle\} \\ \frac{1}{2}\big[f_K(E_i,\mu_K) + f_K(E_j,\mu_K) - 2\sum_{k=1}^{N_K}f_K(E_k,\mu_K)\big]\rho_{i,j} & i\neq j;\ |i\rangle,|j\rangle \in \{|k\rangle\} \\ -\frac{1}{2}\big[\sum_{k\neq i}^{N_K}f_K(E_k,\mu_K)\big]\rho_{i,j} & i\neq j;\ |i\rangle \in \{|k\rangle\}, |j\rangle \notin \{|k\rangle\} \\ -\frac{1}{2}\big[\sum_{k\neq j}^{N_K}f_K(E_k,\mu_K)\big]\rho_{i,j} & i\neq j;\ |i\rangle \notin \{|k\rangle\}, |j\rangle \in \{|k\rangle\} \\ 0 & i\neq j;\ |i\rangle,|j\rangle \notin \{|k\rangle\} \end{cases} \tag{16}$$

The partial trace of the density matrix over the states of lead K appearing in the expression for the non-zero diagonal terms in Eq. (16) provides the instantaneous number of electrons in the finite lead model $\sum_{k=1}^{N_K} \rho_{k,k} = N_K$. Similarly, the sum over Fermi-Dirac weights results in the corresponding equilibrium number of electrons $\sum_{k=1}^{N_K} f_K(E_k, \mu_K) = N_K^{Eq}$. Hence, for $i \in \{|k\rangle\}$ we obtain:

$$\langle i|\mathcal{L}_L|i\rangle = \gamma_K\left[f_K(E_i, \mu_K)N_K - N_K^{Eq}\rho_{i,i}\right] = \gamma_K N_K^{Eq}\left[f_K(E_i, \mu_K)\frac{N_K}{N_K^{Eq}} - \rho_{i,i}\right]. \quad (17)$$

In the limit of infinite lead models the effect of the extended molecule on the leads becomes negligible and the Lindblad operators of Eq. (2) drive the lead occupations toward their equilibrium state such that $\frac{N_K}{N_K^{Eq}} \to 1$ thus we obtain:

$$\langle i|\mathcal{L}_L|i\rangle \xrightarrow[N_K \to \infty]{} \gamma_K N_K^{Eq}\left[f_K(E_i, \mu_K) - \rho_{i,i}\right]. \quad (18)$$

Accordingly, for the off-diagonal terms with $|i\rangle, |j\rangle \in \{|k\rangle\}$ we have:

$$\langle i|\mathcal{L}_K|j\rangle = \frac{\gamma_K N_K^{Eq}}{2}\left[\frac{f_K(E_i, \mu_K) + f_K(E_j, \mu_K)}{N_K^{Eq}} - 2\right]\rho_{i,j}. \quad (19)$$

Since $f_K(E_i, \mu_K)$ and $f_K(E_j, \mu_K)$ are both positive fractions, in the limit of infinite lead models the first term in the square brackets in Eq. (19) vanishes leaving:

$$\langle i|\mathcal{L}_K|j\rangle \xrightarrow[N_K^{Eq} \to \infty]{} -\gamma_K N_K^{Eq}\rho_{i,j}. \quad (20)$$

The remaining non-zero off-diagonal terms of $\mathcal{L}_K$ have either the form:

$$\langle i|\mathcal{L}_K|j\rangle = -\frac{\gamma_K}{2}\left[\sum_{k \neq i}^{N_K} f_K(E_k, \mu_K)\right]\rho_{i,j} = -\frac{\gamma_K}{2}\left[\sum_{k=1}^{N_K} f_K(E_k, \mu_K) - f_K(E_i, \mu_K)\right]\rho_{i,j} =$$
$$-\frac{\gamma_K}{2}\left[N_K^{Eq} - f_K(E_i, \mu_K)\right]\rho_{i,j} = -\frac{\gamma_K N_K^{Eq}}{2}\left[1 - \frac{f_K(E_i, \mu_K)}{N_K^{Eq}}\right]\rho_{i,j} \xrightarrow[N_K^{Eq} \to \infty]{} -\frac{\gamma_K N_K^{Eq}}{2}\rho_{i,j} \quad (21)$$

when $i \in \{|k\rangle\}, j \notin \{|k\rangle\}$ or

$$\langle i|\mathcal{L}_K|j\rangle = -\frac{\gamma_K N_K^{Eq}}{2}\left[1 - \frac{f_K(E_j, \mu_K)}{N_K^{Eq}}\right]\rho_{i,j} \xrightarrow[N_K^{Eq} \to \infty]{} -\frac{\gamma_K N_K^{Eq}}{2}\rho_{i,j}. \quad (22)$$

when $i \notin \{|k\rangle\}, j \in \{|k\rangle\}$ thus yielding the same expression.

Collecting all terms appearing in Eqs. (18), (20), (21), and (22) and substituting in Eq. (16) we finally obtain:

$$\langle i|\mathcal{L}_K|j\rangle = -\gamma_K N_K^{Eq} \begin{cases} [\rho_{i,i} - f_K(E_i, \mu_K)] & i = j; \ |i\rangle, |j\rangle \in \{|k\rangle\} \\ \rho_{i,j} & i \neq j; \ |i\rangle, |j\rangle \in \{|k\rangle\} \\ \frac{1}{2}\rho_{i,j} & i \neq j; \ |i\rangle \in \{|k\rangle\}, |j\rangle \notin \{|k\rangle\} \\ \frac{1}{2}\rho_{i,j} & i \neq j; \ |i\rangle \notin \{|k\rangle\}, |j\rangle \in \{|k\rangle\} \\ 0 & \text{Otherwise} \end{cases} \quad (23)$$

In matrix form the Lindblad superoperator is thus given by:

$$\mathcal{L}\{\boldsymbol{\rho}\} = \mathcal{L}_L\{\boldsymbol{\rho}\} + \mathcal{L}_R\{\boldsymbol{\rho}\} = \begin{pmatrix} -\Gamma_L(\boldsymbol{\rho}_L - \boldsymbol{\rho}_L^0) & -\frac{\Gamma_L}{2}\boldsymbol{\rho}_{L,EM} & -\left(\frac{\Gamma_L}{2} + \frac{\Gamma_R}{2}\right)\boldsymbol{\rho}_{L,R} \\ -\frac{\Gamma_L}{2}\boldsymbol{\rho}_{EM,L} & \mathbf{0}_M & -\frac{\Gamma_R}{2}\boldsymbol{\rho}_{EM,R} \\ -\left(\frac{\Gamma_L}{2} + \frac{\Gamma_R}{2}\right)\boldsymbol{\rho}_{R,L} & -\frac{\Gamma_R}{2}\boldsymbol{\rho}_{R,EM} & -\Gamma_R(\boldsymbol{\rho}_R - \boldsymbol{\rho}_R^0) \end{pmatrix}, (24)$$

where we have defined $\boldsymbol{\rho}_K^0 \equiv \text{diag}\{f_K(E_k, \mu_K)\}$ as a diagonal matrix of dimensions $K \times K$ whose $k_{\text{th}}$ diagonal element is given by $f_K(E_k, \mu_K)$ and $\Gamma_K \equiv \gamma_K N_K^{eq}$. If we further assume that $\Gamma_L = \Gamma_R \equiv \Gamma$ Eq. (24) reduces to:

$$\mathcal{L}\{\boldsymbol{\rho}\} = -\Gamma \begin{pmatrix} (\boldsymbol{\rho}_L - \boldsymbol{\rho}_L^0) & \frac{1}{2}\boldsymbol{\rho}_{L,EM} & \boldsymbol{\rho}_{L,R} \\ \frac{1}{2}\boldsymbol{\rho}_{EM,L} & \mathbf{0}_M & \frac{1}{2}\boldsymbol{\rho}_{EM,R} \\ \boldsymbol{\rho}_{R,L} & \frac{1}{2}\boldsymbol{\rho}_{R,EM} & (\boldsymbol{\rho}_R - \boldsymbol{\rho}_R^0) \end{pmatrix}, \quad (25)$$

which has the exact same structure as the driving term of the driven Liouville von-Neumann equation[90-92] thus completing the derivation.

## **Numerical Demonstration**

In order to demonstrate the correspondence between the Lindblad superoperator defined by Eqs. and (1) and (2) the driving term appearing in Eq. (25) we used both to perform comparative numerical simulations of the electronic transport through a tight-binding atomic chain with increasing lead models size. To this end, we propagate the quantum master equation

$$\dot{\boldsymbol{\rho}}(t) = -\frac{i}{\hbar}[\boldsymbol{H}, \boldsymbol{\rho}] + \mathcal{L}\{\boldsymbol{\rho}\} \quad (26)$$

where the matrix elements of the super-operator $\mathcal{L}$ are given by either Eq. (16) or by Eq. (25). Fig. 2 compares the time-dependent current obtained by the two methods for increasing lead model systems.[97] The model parameters are detailed in the figure caption. Constant driving rates of $\gamma_L = \gamma_R = \gamma = 1 \times 10^{-4} \text{ fs}^{-1}$ are used in all Lindblad equation calculations and the corresponding driving rate used for the driven Liouville von Neumann equation calculations is $\Gamma = 0.015 \text{ fs}^{-1}$. The two values are related via $\Gamma \equiv \gamma N^{eq}$ for the 150 sites lead model system where the average lead equilibrium occupation is $N^{eq} = 150$. As can be seen, the results of the driven Liouville von Neumann calculations quickly converge with lead size such that already for a lead of 150 sites good agreement is obtained with the Landauer steady-state current value. The Lindblad currents, which are consistently smaller than the Driven Liouville von Neumann results, gradually approach the correct steady-state value and the Driven Liouville von Neumann current traces with increasing lead size. This indicates that indeed at sufficiently large lead models the methods become equivalent. Nevertheless, even at a lead size of 300 sites the Lindblad results are not fully converged and deviate from the Landauer value.

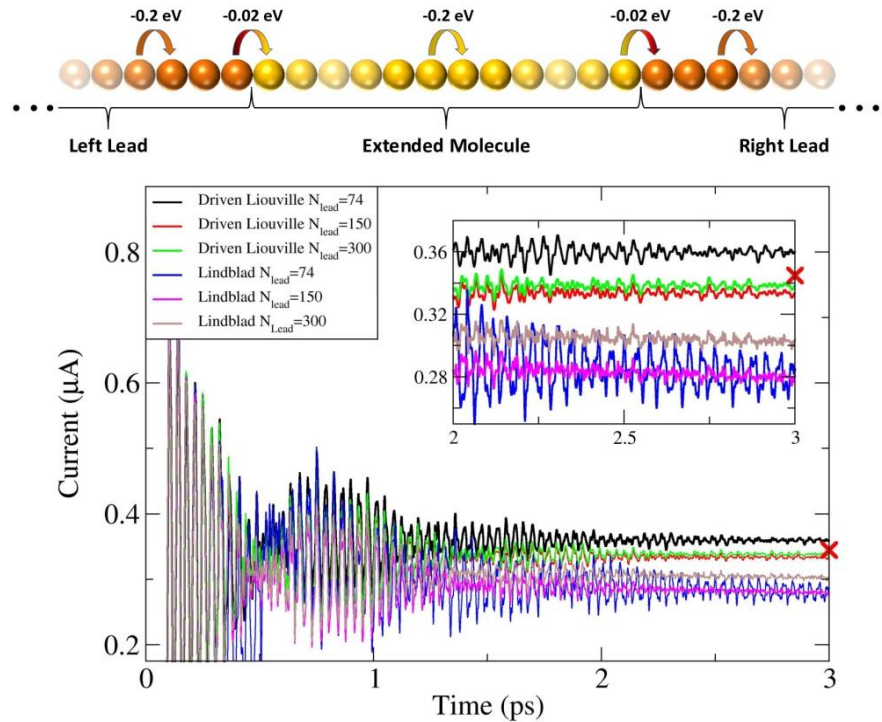

**Figure 2**: Current versus time of a tight-binding atomic chain calculated using the driven Liouville von Neumann equation (black, red, and green) and the Lindbald equation (blue, violet, and brown). Three different lead model sizes with increasing length of 74 (black and blue), 150 (red and violet), and 300 (green and brown) sites are considered. The Landauer stead-state value is presented for reference as the red x-mark. The extended molecule length is 110 sites, all onsite energies are set to zero, the hopping integrals within the left, right, and extended molecule sections are taken to be -0.2 eV and the hopping integral between the two edge sites of the extended molecule and the corresponding edge sites of the leads are taken as -0.02 eV. The bias voltage considered is 0.2 V and the lead electronic temperatures are set to 0 K. The inset provides a zoom in on the steady-state transport region.

To better understand this behavior we analyze the steady-state populations as obtained by the two methods with increasing lead size. In the left panel of Fig. 3 we present the lead and extended molecule state occupations obtained at steady-state using the Driven Liouville von Neumann equation of motion for the three tight-binding chain lead sizes. Notably, all populations are between 0 and 1 indicating that $N$-representability is conserved. Furthermore, the molecular occupations are converged already for the shortest lead model considered of 74 sites while the lead occupations deviations from the target equilibrium distribution reduce with increasing lead size.

The corresponding results for the Lindblad equation are presented in the right panel of Fig. 3. As may be expected all state occupations are positive. Nevertheless, large deviations of the lead steady-state occupations from the target Fermi-Dirac distributions appear. These are characterized by clear violations of the $N$-representability condition leading to occupations that exceed one particle per state in disagreement with Pauli's exclusion principle. The deviations reduce with increasing lead size but are still quite significant even at a lead model size of 300 sites. Accordingly, the extended molecule steady-state occupations, that resemble those obtained by the driven Liouville von Neumann approach, are not fully converged at this lead size. These results further indicate the robustness of the Driven Liouville von Neumann equation for describing time-dependent transport in molecular junctions.

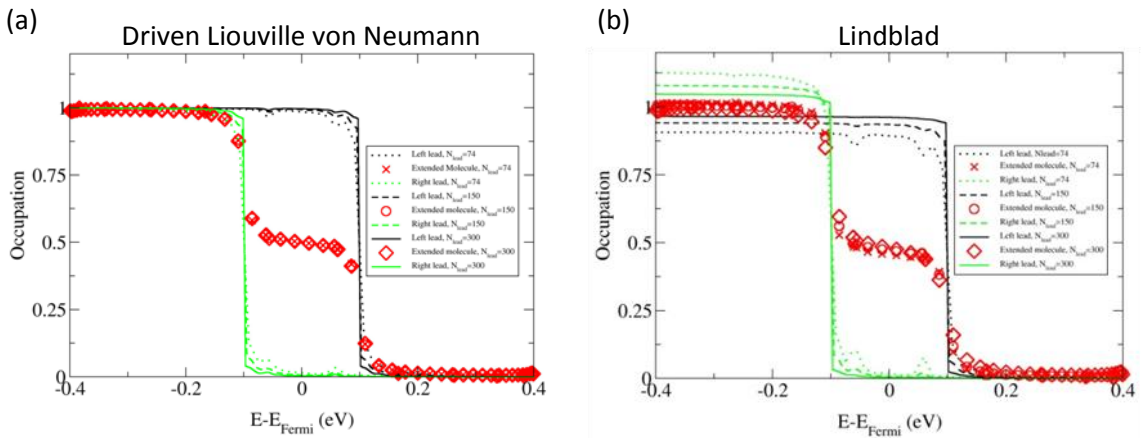

**Figure 3**: Steady-state occupations of the left lead (black), extended molecule (red) and right lead (green) single-particle states as obtained using the driven Liouville von Neumann (left) and the Lindblad (right) master equations for increasing lead sizes of 74 (dotted lines), 150 (dashed lines), and 300 (full lines) sites. The model parameters are detailed in the caption of Fig. 2.

## Summary and conclusions

To summarize, we have shown that in the limit of infinite lead model size the driven Liouville von Neumann equation of motion can be written in Lindblad form. This rationalizes the numerical observation of density matrix positivity conservation and trace preservation exhibited by this equation. Furthermore, it establishes a link between the exact equation of motion, to which the driven Liouville von Neumann equation was shown to be an approximation, and the Linblad operators adopted herein.

## Acknowledgements

O.H. would like to thank Prof. Leeor Kronik, Dr. Yonatan Dubi, and Dr. Felipe Barra for insightful discussions on the subject. Work at TAU was supported by the German-Israeli Foundation under research Grant No. 2291-2259.5/2011, the Israel Science Foundation under Grant No. 1740/13, the Raymond and Beverly Sackler Fund for Convergence Research in Biomedical, Physical and Engineering Sciences, the Lise-Meitner Minerva Center for Computational Quantum Chemistry, and the Center for Nanoscience and Nanotechnology at Tel-Aviv University.